# Effect of morphology and defectiveness of graphene-related materials on the electrical and thermal conductivity of their polymer nanocomposites

S. Colonna[a], O. Monticelli[b], J. Gomez[c], C. Novara[d], G. Saracco[e], A. Fina[a,*]

[a]*Dipartimento di Scienza Applicata e Tecnologia, Politecnico di Torino, 15121 Alessandria, Italy*
[b]*Dipartimento di Chimica e Chimica Industriale, Università di Genova, 16146 Genova, Italy*
[c]*AVANZARE Innovacion Tecnologica S.L., 26370 Navarrete (La Rioja), Spain*
[d]*Dipartimento di Scienza Applicata e Tecnologia, Politecnico di Torino, 10129 Torino, Italy*
[e]*Istituto Italiano di Tecnologia, 10129 Torino, Italy*
*\*Corresponding author: alberto.fina@polito.it*

**Abstract**

In this work, electrically and thermally conductive poly (butylene terephthalate) nanocomposites were prepared by in-situ ring-opening polymerization of cyclic butylene terephthalate (CBT) in presence of a tin-based catalyst. One type of graphite nanoplatelets (GNP) and two different grades of reduced graphene oxide (rGO) were used. Furthermore, high temperature annealing treatment under vacuum at 1700°C was carried out on both RGO to reduce their defectiveness and study the correlation between the electrical/thermal properties of the nanocomposites and the nanoflakes structure/defectiveness. The morphology and quality of the nanomaterials were investigated by means of electron microscopy, x-ray photoelectron spectroscopy, thermogravimetry and Raman spectroscopy. Thermal, mechanical and electrical properties of the nanocomposites were investigated by means of rheology, dynamic mechanical thermal analysis, volumetric resistivity and thermal conductivity measurements. Physical properties of nanocomposites were correlated with the structure and defectiveness of nanoflakes, evidencing a strong dependence of properties on nanoflakes structure and defectiveness. In particular, a significant enhancement of both thermal and electrical conductivities was demonstrated upon the reduction of nanoflakes defectiveness.

1. **Introduction**

The discovery of graphene [1, 2] has attracted the attention of a worldwide scientific community owing to its outstanding electrical, thermal and mechanical properties [3-6]. Unfortunately, despite different synthetic techniques have been developed, including mechanical cleavage of graphite [2], epitaxial growth on SiC [7], chemical vapor deposition [8-10] and liquid phase exfoliation [11, 12], the industrial scale up of graphene, defined as an individual, single-atom-thick sheet of hexagonally arranged $sp^2$-bonded carbons still remains very challenging. On the other hand, the exploitation of graphene in polymer nanocomposites requires large scale production, so that other graphene-related materials (GRM) are typically used, including graphene oxide (GO), reduced graphene oxide (rGO), multilayer graphene (MLG) and graphite nanoplatelets (GNP) [13], which can be easily obtained in quantities suitable for their inclusion in a polymer bulk. Unfortunately, preparation techniques such as chemical reduction of graphene oxide [14, 15], thermal exfoliation and reduction of GO [16], liquid phase exfoliation of graphite [12, 17], electrochemical exfoliation of graphite [18] and other techniques [19] often lead to relatively low quality of the flakes, in terms of average thickness and distribution as well as in chemical defectiveness of the $sp^2$ structure. It is worth noting that in previously reported literature, polymer nanocomposites were prepared with different graphene-related materials, sometimes simply misreferred to as graphene, while a more appropriate nomenclature should be used [13].

The preparation of high quality polymer nanocomposites exploiting graphene-related materials clearly requires a high dispersion degree of the flakes and a good interfacial interaction between the matrix and the nanoflakes, similarly to the case of polymer/CNT nanocomposites [20]. Graphene functionalizations, either covalent and non-covalent, have been demonstrated to be a good solution for the preparation of high-performance polymer nanocomposites [21]. Covalent functionalization can be highly efficient for the improvement of mechanical properties of polymer nanocomposites, directly affecting the stress transfer between polymer matrix and reinforcing particles: however, chemical bonding of organic moieties to an $sp^2$ layer inevitably creates defects ($sp^3$ carbons) which drastically affect electronic, optical and thermal properties of graphene [21, 22]. On the other hand, non-covalent functionalization has a less severe effect on graphene properties, but the reversible adsorption and the weaker interfacial interaction limit its applications in polymer nanocomposites [21].

Thermally conductive polymer nanocomposites are of great interest for those applications where corrosion resistance, easy of processability and lightweight are required. Graphene, owing to its outstanding thermal conductivity, is one of the main candidates for this purpose. Compared to polymer nanocomposites based on carbon nanotubes [20], the use of graphene-related materials appears to be more efficient in terms of thermal conductivity increase, as summarized in Figure 1.

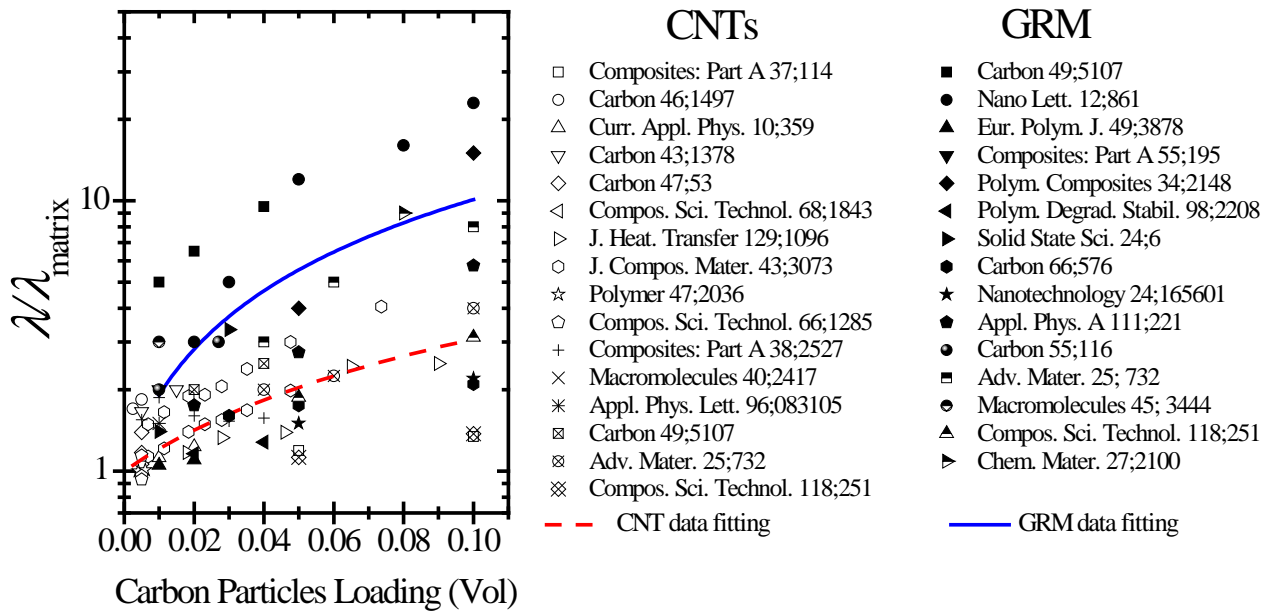

**Figure 1. Normalized thermal conductivity (λ) data of CNT or GRM polymer nanocomposites as function of volume carbon particle loading. Data derived from cited references. Data fitting, reported as a guide for the eye, was performed by calculating the average value at every loading then applying a linear fit**

Despite the obvious scatter of results due to the different preparation procedure, graphene and CNT types, polymer type, measuring technique, etc., the average slope of thermal conductivity increase with the loading of carbon particles is clearly higher for graphene related materials compared to CNT. For both graphene related materials and CNT based nanocomposites, many parameters are recognized to play a crucial role in the improvement of thermal conductivity, including interfacial thermal resistances, nanoparticles quality, dispersion and alignment [20].

Proper dispersion of graphene related materials in polymers often requires energy intensive solvent-assisted processing (e.g. ultrasonication), whereas direct melt processing in polymers remains challenging [21]. Recently, the possibility to obtain an excellent dispersion of GNP, GO and rGO in polymer nanocomposites was reported via in-situ ring-opening polymerization (ROP) of low viscosity cyclic butylene terephthalate (CBT) oligomers into poly-butylene terephthalate (pCBT) during melt blending with carbon nanoparticles [23, 24]. Chen *et al.* [25] reported electric percolation threshold to occur at about 1.6 wt.% of thermally reduced graphene oxide. Noh *et al.* [24], compared different commercially available graphite nanoplatelets with a graphene oxide and a chemically reduced graphene oxide synthetized in a laboratory scale, observing that a higher dispersion degree was obtained when oxidized carbons were present on surface and edges of graphite nanoplatelets. Furthermore, Balogh *et al.* [26] showed a significant reinforcing effect of GNP on the storage modulus (at 25°C the increase was about 85% at 5 wt.% loading, value deduced from plotted data) coupled with a limited improvement in the thermal conductivity of pCBT/graphene nanocomposites. Effects of graphite nanoflakes on the polymerization kinetics of CBT were also reported, in terms of reduction of the average molecular weight [23] and increase of polymerization time [25].

In this work, the preparation of pCBT nanocomposites via ring-opening polymerization in the presence of graphite nanoplatelets or reduced graphene oxide is addressed. Two different types of rGO and one GNP were used with the aim of comparing mechanical, thermal and electrical properties of the nanocomposites prepared with different carbon nanoflakes, with particular attention to their electrical and thermal conductivity. Furthermore, based on the results of our previous work on rGO defectiveness reduction upon high temperature annealing [27], annealed rGO were also used to investigate for the first time the correlations between the structural defectiveness of the rGO and the physical properties of the corresponding nanocomposites.

2. **Experimental**

2.1. **Materials**

Pellets of cyclic butylene terephthalate oligomers [CBT100, Mw = $(220)_n$ *g/mol*, *n* = 2-7, melting point= 130 ÷ 160°C] were purchased from IQ-Holding[1] (Germany). Butyltin chloride dihydroxide catalyst (96%, $m_p$ = 150°C, CAS # 13355-96-9) was purchased from Sigma-Aldrich while acetone (99+%) was purchased from Alfa Aesar.

Three types of graphitic nanoflakes with different surface area were used for this study. GNP (Surface Area = 22 ± 5 $m^2$/g) and RGO (Surface Area = 210 ± 12 $m^2$/g, apparent density 0.002 g/l) were research grades (see below for preparation method) synthetized by AVANZARE (Navarrete, La Rioja, Spain). The second grade of RGO was EXG98 350R, from now named RGO-2, (Surface Area > 300 m$^2$/g, D$_{50}$ < 11 μm by Graphite Kropfmühl (Hauzenberg, Germany).

---

[1] Distributor of products previously commercialized by Cyclics Europe GmbH

## 2.2. Nanoflakes synthesis

Graphite nanoplatelets were prepared using a rapid thermal expansion of intercalated graphite (GIC). The intercalation of graphite with sulfuric acid to obtain graphite-sulphate is a well-known technology described for the first time by Hofmann and Rüdorff [28]. In the present paper, GIC was prepared starting from 500 g of natural graphite flakes (average lateral size ≈ 600 μm) and 5 kg of sulfuric acid were added in a 10 liters glass jacket reactor under continuous stirring at T < 10°C. Then, 200 g of KMnO$_4$ were added to the suspension, keeping the temperature below 10°C. After the complete addition of permanganate, the system was heated up to 50°C and kept at this temperature for 4 hours to allow the completion of the reaction (indicated by the suspension color change to black). At this point the system was cooled to room temperature and the solution was added to about 50 liters of refrigerated H$_2$O, using a peristaltic pump, keeping the temperature lower than 70°C by the external refrigeration in the glass jacket. Hydrogen peroxide (400 g, 30 v.%) was slowly added to remove the excess of MnO$_4^-$, and the suspension was maintained under stirring overnight at room temperature. The solution was then washed in 30 L of agitated 3.3 wt.% HCl solution for 1h. Finally, the solid was filtered, washed with osmotic water, until sulfate test gave a negative result (i.e. adding the solid to a 10 wt.% BaCl$_2$ water solution does not reveal any turbidity), dried first in air and then at 80 °C and, mechanically milled. The solid obtained at this point was named GIC-1. GIC-1 was then introduced in a tubular furnace (N$_2$ atmosphere) at 1000°C for thermal expansion; a worm-like solid was obtained and mechanical milling was required to separate nanoflakes and obtain GNP.

Reduced graphene oxide was produced by thermal reduction of graphene oxide, previously synthetized using a modified Hummers method [29] starting from 250 g of natural graphite (600 μm  average lateral size). The reaction temperature inside the reactor was maintained between 0 and 4°C during oxidant addition (48h; 15.6 Kg H$_2$SO$_4$ 98%; NaNO$_3$, 190 g; KMnO$_4$ 1200 g). Then, the temperature was gradually increased to 20ºC and kept constant for 5 days. A H$_2$O$_2$ solution (50 L H$_2$O; 750 g H$_2$O$_2$ 30%) was used to remove the excess of MnO$_4^-$ over a period of 24 hours. After sedimentation, the solution was washed in a mechanically stirred HCl 4 wt.% solution for 8 h (600:1 washing solution: graphite). The solid was filtered, washed with osmotic water and dried at 80ºC. 100 g of GO were ultrasonicated in isopropanol for two times and placed at reflux overnight. The solid was then removed, filtered and air-dried. This product was later treated in inert atmosphere (Ar) at 1000 °C for 30 sec for the thermal expansion and, then, at 1150 °C for 20 min, leading to the obtainment of a black solid, which is from now on referred to as RGO.

Part of the RGO was annealed, in a closed graphite box, at 1700 °C for 1 hour at 50 Pa in a Pro.Ba. (Italy) vacuum oven with graphite resistors, at the heating and cooling rate of 5 °C/min: this treatment was previously demonstrated sufficient to obtain a significant reduction of defectiveness in the $sp^2$ structure and consequent improvement of heat transfer upon annealing [27]. We will be referring to this annealed material as RGO_1700. The same annealing procedure was applied to RGO-2 and the resulting product is from now on referred to as RGO-2_1700.

### 2.3. Nanocomposite preparation

Nanocomposites were prepared via a 2-step procedure:

1- About 17 g of CBT were partially dissolved in 120 ± 10 ml of acetone for 2 hours under vigorous stirring. Then, the required amount of GNP or rGO was added to the solution and the system underwent a manual mixing for about 5 minutes. The obtained mixture was first dried in a chemical hood for 2h then in an oven at 80°C for 8 hours under vacuum (~$10^1$ mbar) to extract residual acetone and moisture, which could hinder CBT polymerization.

2- The dried mixture was pulverized, loaded into a co-rotating twin screw micro-extruder (DSM Xplore 15, Netherlands) and processed at 100 rpm for 5 minutes; after this period, 0.5 wt.% of butyltin chloride dihydroxide catalyst (with respect to the oligomer amount) was added and the extrusion carried on for other 10 minutes at 100 rpm to complete CBT polymerization into pCBT. The temperature was kept constant for the whole process at 250°C. A nitrogen flux was used to reduce thermoxidative degradation and hydrolysis of polymer.

### 2.4. Characterization

Raman spectra were collected on a Renishaw inVia Reflex (Renishaw PLC, United Kingdom) microRaman spectrophotometer equipped with a cooled charge-coupled device camera directly, on powder deposited on glass slide. Samples were excited with a diode laser source (514.5 nm, 2.41 eV), with a power of 10 mW. The spectral resolution and integration time were 3 cm$^{-1}$ and 10 s, respectively. The deconvolution of D, G and G' peaks was performed by fitting with Lorentzian functions.

X-ray Photoelectron Spectroscopy (XPS) was performed on a VersaProbe5000 Physical Electronics X-ray photoelectron spectrometer with a monochromatic Al source and a hemispherical analyzer. Survey scans as well as high resolution spectra were recorded with a 100 μm spot size. Carbon nanoflakes were fixed on adhesive tape and kept under vacuum overnight to remove volatiles. The deconvolution of XPS peaks was performed with a Voigt function (Gaussian/Lorentzian = 8/2) after Shirley background subtraction.

Morphological characterization of graphene-related materials and pCBT nanocomposites was carried out on a high resolution Field Emission Scanning Electron Microscope (FESEM, ZEISS MERLIN 4248). GNP and rGO, adhered on adhesive tape, were directly observed without any further preparation. pCBT nanocomposites were fractured in liquid nitrogen to avoid plastic deformation, then coated with about 5 nm of Chromium to avoid electrostatic charge accumulation on the surface.

Thermogravimetric analysis (TGA) was performed by placing samples in open alumina pans on a Q500 (TA Instruments, USA), from 50 to 800 °C at the rate of 10 °C min$^{-1}$ with a gas flow of 60 ml min$^{-1}$. The data collected were $T_{max}$ (temperature at maximum rate of weight loss), $T_{onset}$ (the temperature at which the mass lost is 3% of the initial weight) and final residue at 800 °C. TGA on graphene related materials was performed using about 2 mg samples under air flow (oxidative atmosphere).

Differential scanning calorimetry (DSC) was carried out on a Q20 (TA Instruments, USA) with a heating rate of 10 °C min$^{-1}$ in the temperature range 25 to 240 °C. The method consisted on a first heating cycle, performed to erase the thermal history of the material, a cooling step, to study the crystallization of pCBT and a last heating step to evaluate the melting temperature of materials. Crystallinity was calculated as the ratio between the integrated value for heat of melting of the sample and the heat of melting of 100% crystalline poly (butylene terephthalate), i.e. 140 J/g [30], and normalized in nanocomposites taking into account of the effective polymer fraction in the sample.

Dynamic-mechanical thermal analysis (DMTA) was implemented on a Q800 (TA Instruments, USA) with tension film clamp on ca. 1 mm thick specimens. The experimental conditions were: temperature range from 30 to 190 °C, heating rate of 3 °C min$^{-1}$, frequency equals to 1 Hz and 0.05% of oscillation amplitude in strain-controlled mode.

Rheological properties of pCBT/GNP nanocomposites were measured on a strain-controlled rheometer (Advanced Rheometric Expansion System, ARES, TA Instruments, USA) with parallel-plate geometry (diameter of 25 mm). The instrument was equipped with a convection oven to control the temperature. Before the measurements, dried nanocomposites were pressed at 250 °C into disks with thickness and diameter of 1 and 25 mm, respectively. Specimens were further dried at 80 °C in vacuum for 8 h before the measurements to avoid water absorption. Oscillatory frequency sweeps ranging from 0.1 to 100 rad/s with a fixed strain (chosen and selected for each sample in order to fall in the linear viscoelastic region) were performed in air at 240 °C, to investigate the viscosity of the nanocomposites. After the sample loading, an approximate 5 min equilibrium time was applied prior to each frequency sweep.

Electrical resistivity (volumetric) was evaluated with a homemade apparatus on the same specimens as those for the rheological tests. The apparatus for the measurement is composed by:

- A tension and direct current regulated power supply (PR18-1.2A of Kenwood, Japan);
- A numeral table multimeter (8845A of Fluke, Everette/USA) equipped with a digital filter in order to reduce the noise of the measure;
- A palm-sized multimeter (87V of Fluke, Everette/USA);
- Two homemade electrodes: both made of brass. One is a cylinder (18,5mm diameter, 55mm height) and the other is a plate (100mm side, 3mm thickness). Every electrode has a hole for the connection and a wire equipped with a 4mm banana plug.

The measurement system works with the multimeter method. The power supply is time to time regulated in current or in voltage to have accurate measurement by both the multimeters, limiting, however, the power dissipated on the specimen. The resistivity value was calculated with the following formula:

$$\rho = \frac{V}{I} \cdot \frac{S}{l} \ [\Omega \cdot m] \tag{1}$$

where S and l are the specimen surface and thickness, respectively, while V is the voltage and I the electric current, both read by the measurement apparatus.

Isotropic thermal conductivity tests were performed on a TPS 2500S (Hot Disk AB, Sweden) with a Kapton sensor (radius 3.189 mm). Before the measurements, dried nanocomposites were pressed at 250 °C into disks with thickness and diameter of about 4 and 15 mm, respectively. Specimens were further stored in a constant climate chamber (Binder KBF 240, Germany) at 23.0 ± 0.1 °C and 50.0 ± 0.1 %R.H. for at least 48 h before the measurements. Sensor was inserted between two specimens of the materials under examination and the overall system was put in a container dipped into a silicon oil bath (Haake A40, Thermo Scientific Inc., USA) equipped with a temperature controller (Haake AC200, Thermo Scientific Inc., USA). All the materials were tested at 23.00 ± 0.01 °C.

## 3. Results and discussion

### 3.1. Nanoflakes characterization

In this work, the effect of GNP and rGO, on mechanical, electrical and thermal properties of their pCBT-based nanocomposites was studied. With this aim, GNP and rGO were thoroughly characterized for their morphology and structural defects.

In Figure 2, the morphology of the as received/as prepared powders of different carbon nanoflakes used for this study is shown. GNP (Figure 2a) consists of aggregated flat nanosheets with lateral size ranging from few hundred nanometers to several micrometers and thickness estimated in the range of 10-20 nm. On the other hand, powders of RGO and RGO-2 (Figure 2b,d) consist of accordion-like organization of a few nanometers thick and highly wrinkled layers with an irregular shape and lateral size estimated in the range of a few micrometers, with minor differences observable between RGO and RGO-2. The morphology of both reduced graphene oxide grades was not significantly altered by the treatment at 1700°C, even if the accordion-like structure appears slightly more expanded after annealing (Figure 2c,e). Additional micrographs for the different nanoflakes are reported in Supporting Info (Figure S1).

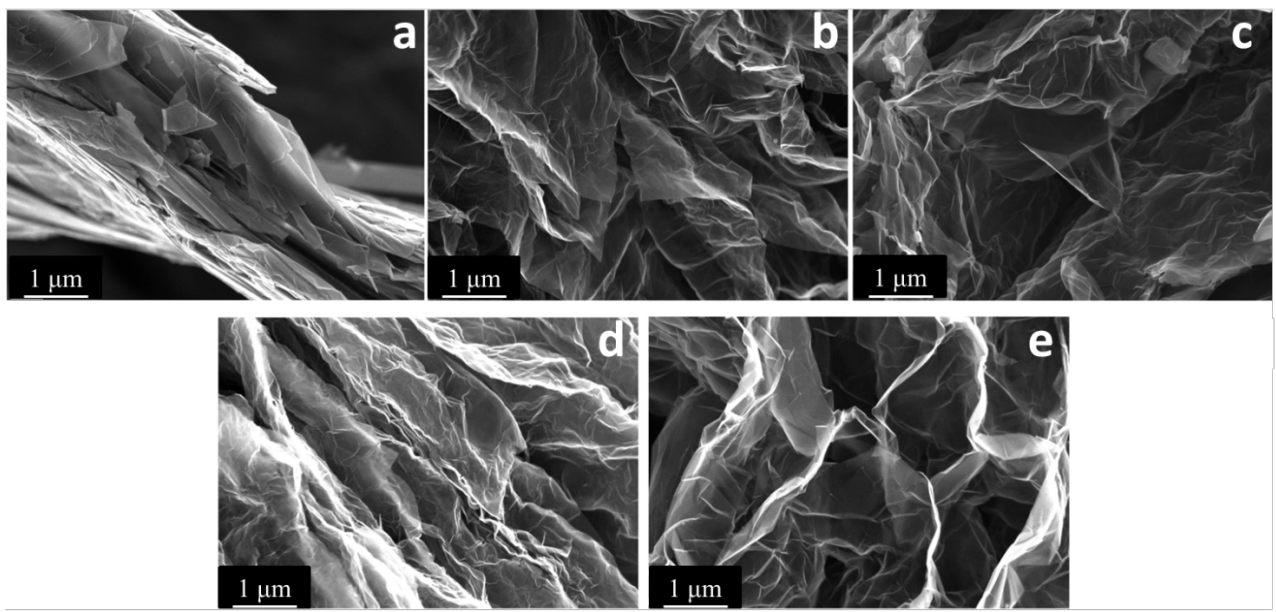

**Figure 2. FESEM micrographs for a) GNP, b) RGO, c) RGO_1700, d) RGO-2 and e) RGO-2_1700**

Raman spectra were collected to evaluate the defectiveness of nanoflakes. First- and second-order Raman spectra are reported, normalized with respect to the G peak (ca. 1580 cm$^{-1}$), in Figure 3. The first-order Raman spectrum for GNP exhibits a tiny signal at ~1350 cm$^{-1}$ (defect-related D band) and a strong band at ~1578 cm$^{-1}$ (G band). Thus, the $I_D/I_G$ ratio (Table 1) is very small, evidencing very limited defectiveness of this GNP. The second-order band at higher Raman shift is the convolution of two main peaks (G'$_1$ and G'$_2$) located at ~2690 cm$^{-1}$ and ~2725 cm$^{-1}$, respectively, which are typical for graphitic materials constituted by more than 5 graphene layers [31].

For both rGO, a radically different Raman spectrum was observed, mainly in terms of a strong D band at ~1345 cm$^{-1}$ with intensity comparable to the G band. It is worth noting that deconvolution of the first-order Raman fingerprint for RGO-2 requires the addition of a further band located at ~1521 cm$^{-1}$ which was previously ascribed to highly disordered areas [32] or to amorphous carbon [33]. The second-order Raman spectrum was characterized by a very weak G' mode, consisting of a wide band at ~2700 cm$^{-1}$ and a reliable fitting with multiple peaks was not feasible. In conclusion, based on the $I_D/I_G$ ratio (Table 1) and the overall features of the Raman spectra, a high concentration of defects was clearly evidenced in the sp$^2$ structure for both RGO and RGO-2 [33, 34]. After thermal annealing at high temperature dramatic changes in Raman spectra were observed for both RGO grades. Indeed,

RGO_1700 and RGO-2_1700 exhibit very similar Raman spectra, with the same $I_D/I_G$ ratio (0.11). In fact, the G band shrinks and increases its intensity upon high temperature treatment while the D band becomes very weak. Furthermore, the second-order spectrum for both rGO shows a narrower and more intense G' peak at ~2705 cm$^{-1}$.

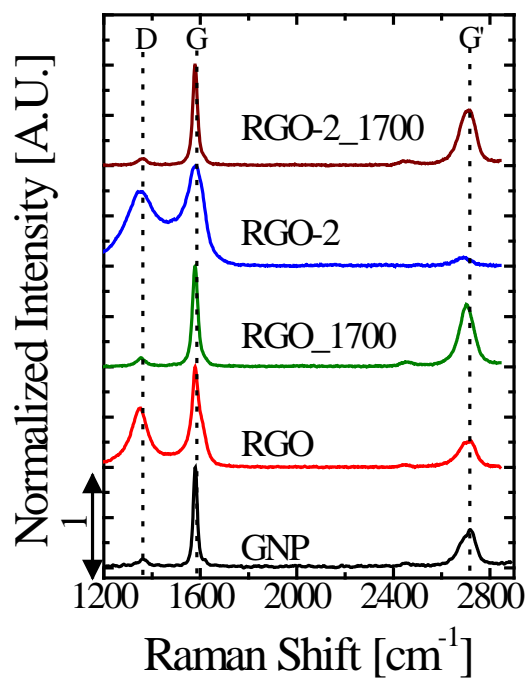

**Figure 3. Representative Raman spectra for the different nanoflakes**

XPS was used to study the chemical compositions of flakes before and after annealing. The oxygen content for the different nanoflakes, calculated by integration of survey scan peaks (Table 1) were measured at 1.8, 3.2 and 7.0 at.% for GNP, RGO and RGO-2, respectively. The annealing treatment drastically decrease the content of oxidized groups down to 0.4 and 1.1 at.% for RGO_1700 and RGO-2_1700, respectively. A closer insight on the chemical structures of graphite nanoflakes was carried out by narrow scans, and following peak deconvolution, on $C_{1s}$ (B.E. ≈ 285 eV) and $O_{1s}$ (B.E. ≈ 530 eV) signals. Detailed results of carbon and oxygen bands deconvolution are reported and discussed in the supporting information (Figure S1).

Thermal stability of nanoflakes was evaluated by TGA in air (see Supporting Info for mass vs. temperature plot, Figure S2), to indirectly investigate their structural features, knowing that the onset decomposition temperature can be qualitatively related to the size and the defectiveness of graphene-related materials [35]. $T_{onset}$ (Table 1) for GNP was measured at 632°C, whereas significantly lower values were obtained for RGO (558°C) and RGO-2 (471°C). Furthermore, a dramatic increase in $T_{onset}$ was observed after high temperature annealing, up to 750 °C for RGO_1700 and 671 °C for RGO-2_1700. These results provide additional evidence for the reduction of rGO defectiveness upon high temperature annealing, in agreement with Raman and XPS results described above.

**Table 1. Quantification of defectiveness of nanoflakes by Raman, thermal stability (TGA) and oxygen content (XPS).**

| Nanoflake | $I_D/I_G$ | $T_{Onset}$ TGA [°C] | $O_{1s}$ [at.%] |
|---|---|---|---|
| GNP | 0.16 ± 0.02 | 632 ± 12 | 1.8 ± 0.1 |
| RGO | 0.88 ± 0.33 | 558 ± 10 | 3.2 ± 0.1 |
| RGO_1700 | 0.11 ± 0.02 | 750 ± 21 | 0.4 ± 0.1 |
| RGO-2 | 0.80 ± 0.01 | 471 ± 16 | 7.0 ± 0.2 |
| RGO-2_1700 | 0.11 ± 0.02 | 671 ± 31 | 1.1 ± 0.1 |

The different structural features of graphene nanoplatelets and reduced graphene oxides, either pristine and after high temperature annealing, make these materials an ideal set for the systematic study of how nanoflakes properties may affect the performance of the corresponding polymer nanocomposites. In the following section, the physical properties of such nanocomposites are discussed in details.

### 3.2. Nanocomposites characterization

The polymerization of CBT into pCBT was monitored by differential scanning calorimeter. No traces of the characteristic melting peaks of CBT were observed after their extrusion in the presence of the tin catalyst (Figure S3), evidencing for complete conversion of CBT oligomers into a polymer. The degree of crystallinity of pCBT was calculated equal to 41.7%. The addition of 5 wt.% of nanoflakes induced a slight reduction in the crystallinity degree of pCBT with values ranging between 34.4% and 40.9% (see Table S1 for further details).

The morphologies of pCBT nanocomposites were studied by electron microscopy: representative micrographs of pCBT + 5 wt.% nanoflakes are reported in Figure 4. Both nanocomposites containing GNP and rGO exhibit a good distribution of nanoflakes with some differences in the morphology. In pCBT + 5% GNP (Figure 4a) relatively large and thick nanoflakes are observed, reflecting the morphology of the starting graphite nanoplatelets (Figure 2a). In nanocomposites containing RGO (Figure 4b) and RGO-2 (Figure 4d) smaller and thinner flakes are observed, suggesting separation and dispersion of the nanometric layers from accordion-like structure. Distribution and dispersion of nanoflakes does not seem to be affected by the thermal treatment as no significant differences are observed between the morphologies of nanocomposites containing pristine (Figure 4b,d) and annealed rGO (Figure 4c,e).

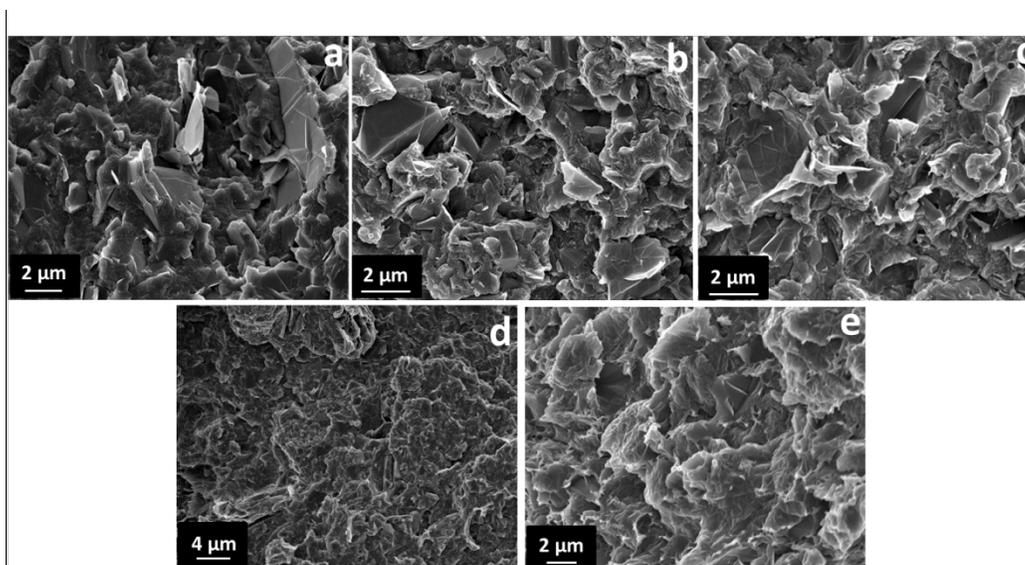

**Figure 4. FESEM pictures of pCBT nanocomposites containing 5 wt.% of a) GNP, b) RGO, c) RGO_1700, d) RGO-2 and e) RGO-2_1700.**

To further study the dispersion of nanoflakes in the polymer and their organization into a percolating network, linear viscoelasticity in the molten state was studied performing dynamic frequency sweep test. Indeed, elastic modulus (G') and complex viscosity η* are well known to be sensitive to the filler content and dispersion [36], thus providing indirect information on the nanoparticles organization in the polymer bulk. G' and η* plots for pCBT + 5% nanoflakes are reported in Figure 5 as a function of deformation frequency. Pure pCBT exhibits the classical behavior of polymers in linear regime, showing G' decrease as the frequency decreases while η* is approximately constant in the whole frequency range (Figure 5). On the contrary, for all the nanocomposites, G' exhibits a weak dependency on the frequency in the whole frequencies range, evidencing the formation of a solid-like network [37, 38], *i.e.* a well-organized percolated structure of the nanoflakes. A further evidence for the high percolation degree of nanoflakes in the nanocomposites is provided by the strong dependence of the complex viscosity with frequency [36], extending over four decades (Figure 5b). Comparing the elastic modulus and viscosity plots for the different nanocomposites, significant differences can be observed. In the case of GNP, clearly lower values for both G' and η* were observed compared with rGO. This reflects the different size of dispersed particles, evidencing for a relatively loose yet percolating network structure. Among nanocomposites containing reduced graphene oxides, differences are clearly visible between pCBT + 5% RGO and pCBT + 5% RGO-2, the latter evidencing a significantly more organized percolation network. However, when using annealed rGO, both nanocomposites exhibited very similar modulus and viscosity plots.

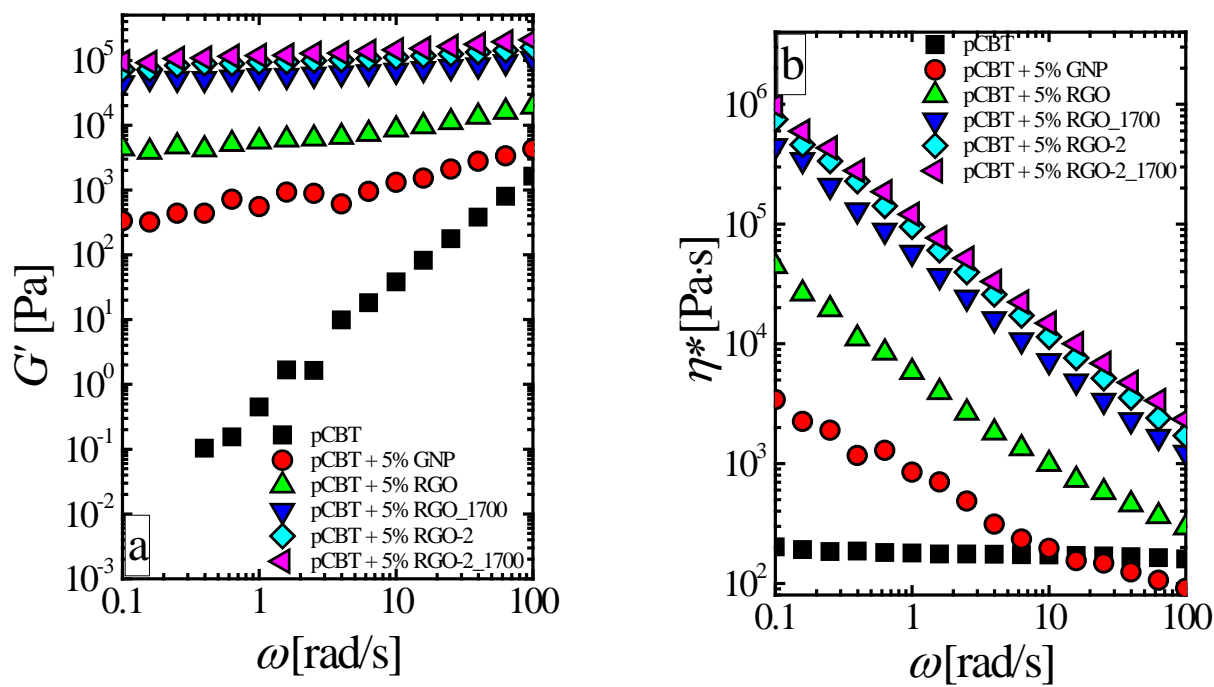

**Figure 5.** Dynamic frequency sweep test at 240°C for pCBT and its nanocomposites. (a) G' and (b) complex viscosity as a function of the angular frequency

The organization of nanoflake into a percolated network and its effect on viscoelastic properties of the nanocomposites was further studied in the solid state by dynamo-mechanic analysis. DMTA was used to evaluate the effect of the different nanoflakes on either storage modulus and glass transition temperature of pCBT and its nanocomposites (Figure 6). The inclusion of nanoflakes was found to strongly increase the storage modulus over the whole temperature range explored, which is consistent with the formation of a stiff network of nanoflakes, with limited differences between GNP and the different rGO. The temperature for the main relaxation of the polymer, measured as the peak of Tanδ plot, is generally increased and /or broadened (Table S2) by the presence of GNP or rGO, compared to the reference pCBT, suggesting confinement of the polymer chains induced by the presence of dispersed nanoflakes.

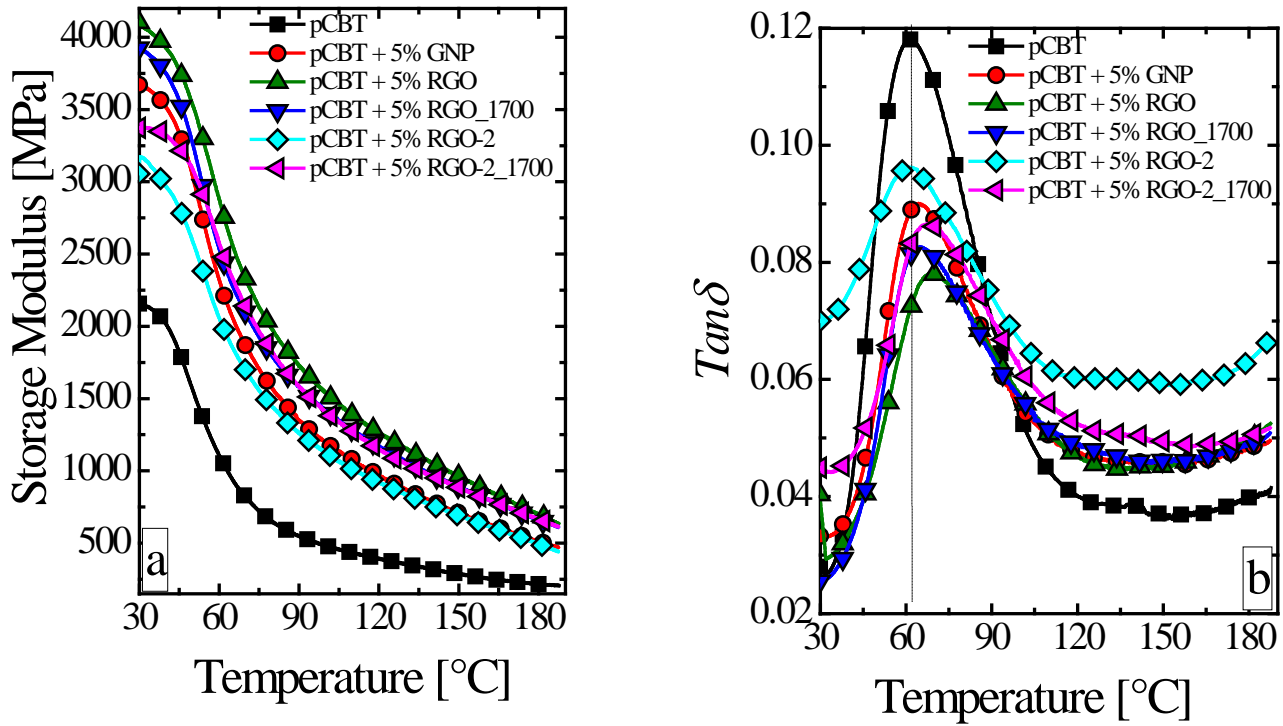

**Figure 6. (a) Storage modulus and (b) tan delta plots measured by DMTA**

While pure pCBT shows a very low electrical conductivity, in the range of $10^{-13}$ S/m which is typical for insulating materials [24, 25], the formation of a percolation network with conductive nanoflakes is clearly expected to result in an electrically conductive material. Indeed, graphene and graphene related materials are good candidates for the improvement of electrical properties of polymeric materials, provided a good dispersion is obtained and the quality of the nanoflakes is sufficient to maintain a high charge mobility [39]. A sharp transition from insulating to conductive materials is typically associated to the percolation threshold, which clearly depends on both the dispersion degree and the nanoflakes aspect ratio: for pCBT mixed with RGO and GNP values ranging between 1.6 wt.% [25] and 5 wt.% [24], respectively.

Electrical conductivity results for pCBT nanocomposites in this paper (Figure 7, Table 2) clearly confirmed all nanocomposites prepared at 5 wt.% loading are above the percolation threshold, with conductivity values in the range of $10^{-5}$ S/m for GNP and between $10^{-4}$ S/m and $10^{-2}$ S/m with the different rGO. The difference between GNP and rGO has to be ascribed to the lower density of the percolation network obtained with graphite nanoplatelets, in agreement with rheology results discussed above. Furthermore, large differences were observed between nanocomposites containing different grades of rGO. In particular, it is worth highlighting that both RGO-1700 and RGO-2_1700 are more effective in improving electrical properties, compared to the corresponding nanocomposites containing pristine RGO and RGO-2. Taking into account the minor differences in dispersion of annealed *vs.* pristine rGO described above on the basis of electron microscopy, rheology and viscoelastic properties, the electrical conductivity results here reported evidence the strong effect of the reduction of nanoflakes defectiveness on their intrinsic electrical conductivity and, in turn, on the electrical conductivity of related polymer nanocomposites.

Further enhanced electrical conductivity was progressively obtained with increasing the nanoflake loading, as summarized in Figure 7. However, it is worth mentioning that the loading of nanoflakes which can be included in pCBT by melt blending is limited by the viscosity of the nanocomposite obtained. While in the case of GNP the increase of viscosity was limited and loading up to 30 wt.% did not cause processing problems, extremely high viscosities were obtained during polymerization at 10 wt.% loading of the different rGO and preparation of pCBT + 10% RGO-2 was not possible in the conditions used for the other preparations. Even with these limitations, the analysis of conductivity results at higher loading clearly confirms that efficiency in electrical conductivity enhancement is maximum for thermally annealed RGO and minimum for GNP. Electrical conductivity values in the range of $10^{-1}$ S/m were obtained for pCBT + 10% RGO_1700 (0.09 ± 0.01 S/m) as well as for pCBT + 30% GNP (0.19 ± 0.004 S/m), the different loading to obtain similar electrical performance evidencing for superior properties of low defectiveness RGO.

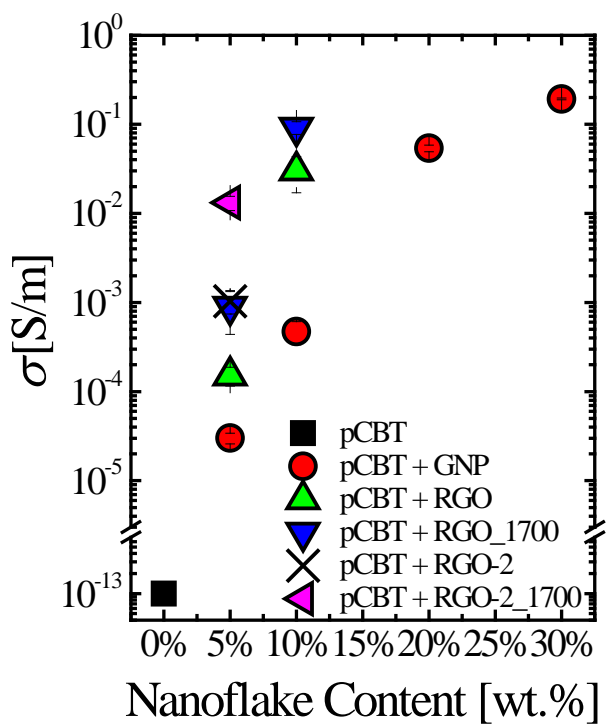

**Figure 7.** Electrical conductivity vs. nanoflake content. The value here reported for pure pCBT was taken from ref. [24]

Bulk thermal conductivity results for pCBT and its nanocomposites with GNP and rGO as a function of filler content are reported in Figure 8 and Table 2. Addition of 5 wt.% of GNP or pristine rGO led to thermal conductivity values, in the range of 0.5 W/(m·K), that is corresponding to about twice the conductivity of pCBT [0.240 ± 0.003 W/(m·K)]. It is well known that this increase is clearly not comparable with the jump in electrical conductivity associated to the percolation threshold. Two different reasons are generally agreed upon as an explanation for this feature. On the one hand, the thermal conductivity ratio between carbon nanoflakes and polymers is in the range of $10^3$-$10^4$, thus much lower than for electrical conductivity (ratio of $10^{12}$-$10^{15}$). On the other hand, the simple physical contact between two particles (mechanical percolation) may be sufficient to allow an electron to hop between particles close enough, but it is indeed insufficient to allow efficient phonon transfer [20]. It is worth noting that thermal conductivity of pCBT + 5% GNP is slightly higher than for both pCBT + 5% RGO and pCBT + 5% RGO-2. This result may appear surprising based on the particles dispersion and electrical conductivities described above, but it can be explained taking into account the high degree of defectiveness in rGO, which was previously demonstrated to strongly affect their intrinsic thermal conductivity [27]. Indeed, when annealed nanoflakes were used, the value of thermal conductivity obtained were 0.890 ± 0.009 and 0.995 ± 0.003 W/(m·K) for pCBT + 5% RGO_1700 and pCBT + 5% RGO-2_1700, respectively, *i.e.* about twice the value obtained for untreated rGO at the same loading. This dramatic increase is clearly related to the rGO structural evolution upon high temperature annealing, as described in the first part of this paper, and evidences experimentally for the first time the correlation between the defectiveness of the rGO nanoflakes and the thermal conductivity of the relevant nanocomposites. Nanocomposites with 10 wt.% of nanoflakes exhibit higher thermal conductivities with a maximum value of 1.772 ± 0.003 W/(m·K) for RGO_1700, *i.e.* about three times the conductivity of pCBT + 10% RGO, thus further confirming the effect of rGO thermal annealing on nanocomposites conductivity. At 10% loading, nanocomposites containing GNP displayed a thermal conductivity between that of RGO and RGO_1700 at the same loading. However, while maximum rGO content is limited by its difficult processability, nanocomposites with higher loading of GNP can be prepared, taking advantage of its moderate effect on melt viscosity, leading to further thermal conductivity increase with the amount of GNP, up to 2.49 ± 0.02 W/(m·K) at 30 wt.% loading. About equivalent thermal properties were found for pCBT + 10% RGO_1700 and pCBT + 20% GNP, [1.772 W/(m·K) and 1.827 W/(m·K), respectively], which would suggest the two materials can be considered alternatives in thermal management applications. However, from a practical point of view, the choice between nanocomposites embedding of GNP at high loading or rGO at lower loading is a matter of other properties beyond the bare values of thermal or electrical conductivities, including for instance density of the material, brittleness and impact resistance as well as processability, recyclability and cost. Furthermore, despite the clear advantages associated with rGO annealing at high temperature, the additional energy input for the additional process has to be taken into account when considering the transfer of the present results to an industrial application. Therefore, while thermal annealing remains a powerful method to boost thermal properties, further efforts remain necessary for the development of efficient large scale production of low defectiveness carbon nanoflakes.

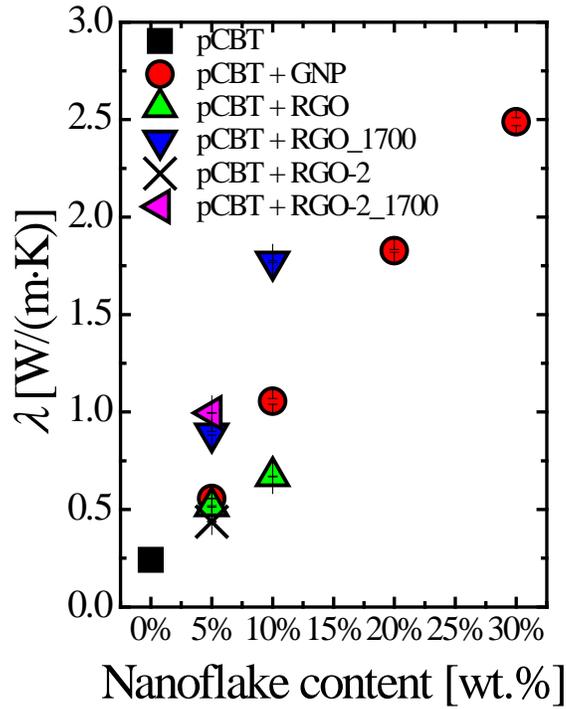

**Figure 8. Thermal conductivity vs. nanoflake content**

**Table 2. Electrical and thermal conductivity data for pCBT nanocomposites**

| | Electrical conductivity [S/m] | | | |
| | Thermal conductivity [W/(m·K)] | | | |
| Nanoflake Content | 5 wt.% | 10 wt.% | 20 wt.% | 30 wt.% |
|---|---|---|---|---|
| pCBT + GNP | $\sigma = (3.0 \pm 0.4)$ E-5 | $\sigma = (4.7 \pm 1.4)$ E-4 | $\sigma = (5.4 \pm 0.5)$ E-2 | $\sigma = (1.9 \pm 0.04)$ E-1 |
| | $\lambda = 0.556 \pm 0.001$ | $\lambda = 1.005 \pm 0.015$ | $\lambda = 1.827 \pm 0.007$ | $\lambda = 2.489 \pm 0.020$ |
| pCBT + RGO | $\sigma = (1.5 \pm 0.4)$ E-4 | $\sigma = (3.0 \pm 1.3)$ E-2 | - | - |
| | $\lambda = 0.515 \pm 0.004$ | $\lambda = 0.669 \pm 0.002$ | | |
| pCBT + RGO_1700 | $\sigma = (9.0 \pm 4.6)$ E-4 | $\sigma = (9.2 \pm 1.5)$ E-2 | - | - |
| | $\lambda = 0.890 \pm 0.009$ | $\lambda = 1.772 \pm 0.005$ | | |
| pCBT + RGO-2 | $\sigma = (1.0 \pm 0.3)$ E-3 | - | - | - |
| | $\lambda = 0.437 \pm 0.003$ | | | |
| pCBT + ROG-2_1700 | $\sigma = (1.3 \pm 0.2)$ E-2 | - | - | - |
| | $\lambda = 0.995 \pm 0.001$ | | | |

## 4. Conclusions

In this work, pCBT nanocomposites were prepared by in-situ ring opening polymerization of CBT in the presence of graphite nanoplatelets or reduced graphene oxide. rGO obtained from different sources were used, both as received and after high temperature annealing to investigate correlation between the morphological/chemical features of carbon nanoflakes and the physical properties of their nanocomposites.

Melt mixing process in low viscosity CBT and subsequent polymerization in extrusion allowed good dispersion of nanoflakes in the polymer, as proven by electron microscopy and rheological analysis, which clearly provided evidences for a highly percolated structure. Significant differences were observed between nanocomposites containing GNP and rGO, in terms of denser percolation network (i.e. higher viscosity) obtained with thinner and smaller rGO nanoflakes compared to larger graphite nanoplatelets.

Electrical and thermal conductivity results showed interesting differences between nanocomposites with GNP and those embedding rGO, as well as between pristine and annealed rGO. The electrical conductivity results directly reflect the properties of the percolation network extrapolated by rheological analysis, the conductivity with rGO being significantly higher (up to two order of magnitude for RGO-2) compared to GNP. High temperature annealing of rGO further enhanced the electrical conductivity, leading to best result in the range of $10^{-1}$ S/m at 10wt.%, which appears to be related to an improvement in charge mobility on the rGO rather than to differences in the percolation network obtained. Thermal conductivity enhancement was also found very different in the presence of GNP, rGO or annealed rGO. Comparison between nanocomposites with GNP and pristine rGO showed better thermal conductivity for GNP, which is in contrast with the electrical and rheological behavior. This result clearly evidence that the bare presence of a well-organized percolating network is not sufficient to obtain a high thermal conductivity of the composite and can only be explained taking into account the strong dependency of the thermal conductivity of the carbon nanoflakes as a function of defectiveness. In nanocomposites containing rGO, nanoflakes defectiveness (oxidized carbons and other defects in the $sp^2$ structure) leads to a drop in their conductivity and, despite a well-organized percolation network was proved, the thermal conductivity performance is lower than for GNP-based nanocomposite, in which the percolation network is looser but nanoplatelets are significantly less defective. When comparing nanocomposites containing annealed vs. pristine rGO, a two- to three-fold increase in thermal conductivity was observed upon high temperature annealing of the rGO. This dramatic increase is clearly related to the reduction in rGO defectiveness rather than to differences in the percolation network. These results provide for the first time, to the best of the author's knowledge, experimental evidences of the correlation between the defectiveness of the rGO and the thermal conductivity of the relevant nanocomposites.

**Acknowledgements**

The research leading to these results has received funding from the European Union Seventh Framework Programme under grant agreement n°604391 Graphene Flagship. This work has received funding from the European Research Council (ERC) under the European Union's Horizon 2020 research and innovation programme  grant agreement 639495 — INTHERM — ERC-2014-STG. Funding from Graphene@PoliTo initiative of the Politecnico di Torino  is also acknowledged.

The authors gratefully acknowledge Dr. Mauro Raimondo for FESEM observations, Salvatore Guastella for XPS analyses and Fausto Franchini for electrical conductivity measurements. Prof. Matteo Pavese and prof. Fabrizio Giorgis are also gratefully acknowledged for high temperature annealing treatments and support in Raman interpretation, respectively.

**Authors contributions**

A. Fina conceived the experiments and coordinated the project, S. Colonna carried out the entire preparation of nanocomposites and most of the characterizations reported in this paper. O. Monticelli contributed to ring opening polymerization design and characterization. J. Gomez prepared GNP and RGO.  C. Novara carried out Raman measurements. G. Saracco contributed to the discussion of the results. Manuscript was written by S. Colonna and A. Fina.

Supporting Information

# Effect of morphology and defectiveness of graphene related materials on the electrical and thermal conductivity of their polymer nanocomposites

S. Colonna[a], O. Monticelli[b], J. Gomez[c], C. Novara[d], G. Saracco[e], A. Fina[a,*]

[a]*Dipartimento di Scienza Applicata e Tecnologia, Politecnico di Torino, 15121 Alessandria, Italy*
[b]*Dipartimento di Chimica e Chimica Industriale, Università di Genova, Via Dodecaneso, 31, 16146 Genova, Italy*
[c]*AVANZARE Innovacion Tecnologica S.L., 26370 Navarrete (La Rioja), Spain*
[d]*Dipartimento di Scienza Applicata e Tecnologia, Politecnico di Torino, 10129 Torino, Italy*
[e]*Istituto Italiano di Tecnologia, 10129 Torino, Italy*

*Corresponding author: alberto.fina@polito.it*

**Scanning electron microscopy**

Additional FESEM micrographs for graphite nanoplatelets and reduced graphene oxide are reported in Figure S1

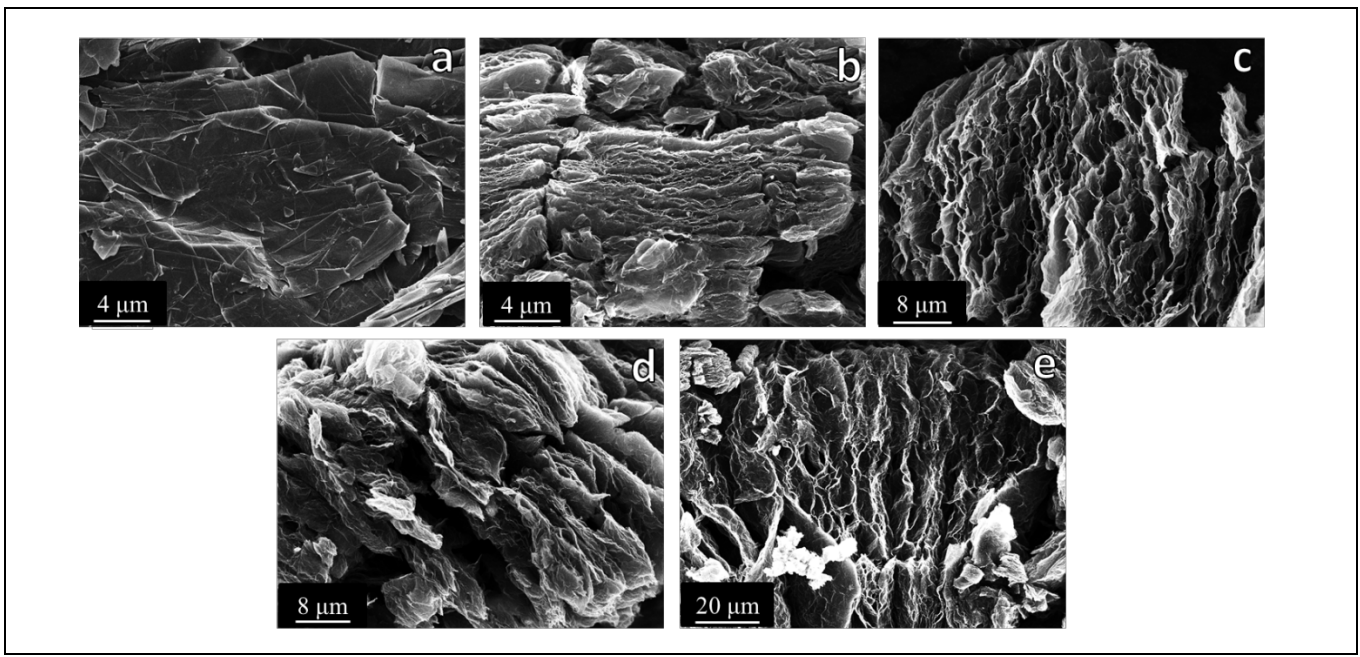

**Figure S1**. Additional FESEM micrographs for a) GNP, b) RGO, c) RGO_1700, d) RGO-2 and e) RGO-2_1700

**X-ray photoelectron spectroscopy**

For all the graphitic nanoflakes, the C1s region shows an intense anisotropic peak with maximum centered at about 284.5 eV and a long tail, up to ~295 eV, related to overlapping of several peaks, whit a shape which is typical for reduced graphene oxide. The peak located at a binding energy (B.E.) of 284.5 eV is assigned to $sp^2$ C-C carbon while chemical shift of 1.5, 2.5 and 4.0 eV are typically assigned to C-O, C=O and COOH functional groups, respectively[1, 2]. However, it is worth noting that in literature the assignment of the different peaks is often subjective and controversial[3, 4]. In this work, fitting of C1s peaks was carried out fixing the position of $sp^2$ C-C (284.5 eV) and π-π* shake-up (291.3 eV) transitions while leaving two additional positions for optimal peaks fitting. Oxygen 1s assignment was reported to be less controversial in literature and various authors attributed binding energy values of ~533.0 and ~531.0 eV for single-bonded and double-bonded (C=O, O=C-OH) oxygen[1, 5, 6], respectively, even if some authors further distinguish between the B.E. of the different chemical bonds[1]. Moreover, it is worth noting that O1s photoelectron kinetic energies are lower than those of C1s, i.e. O1s spectra are slightly more surface sensitive.

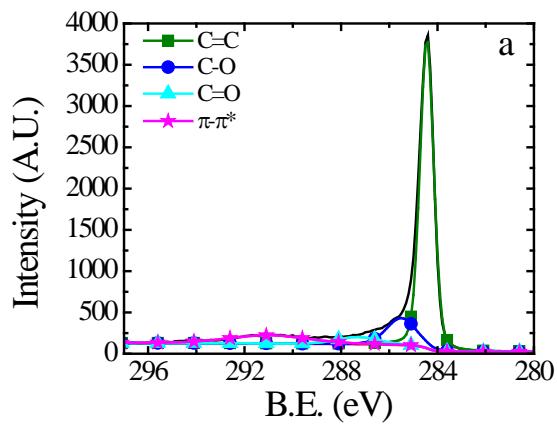
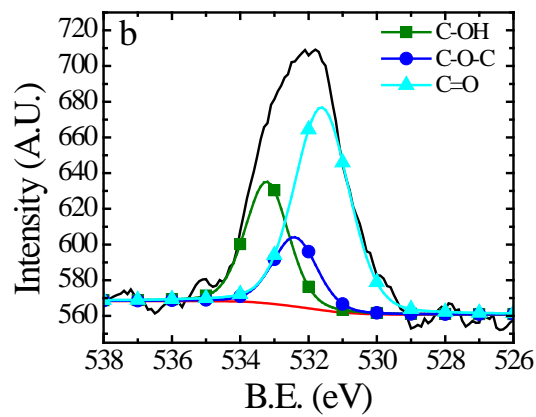
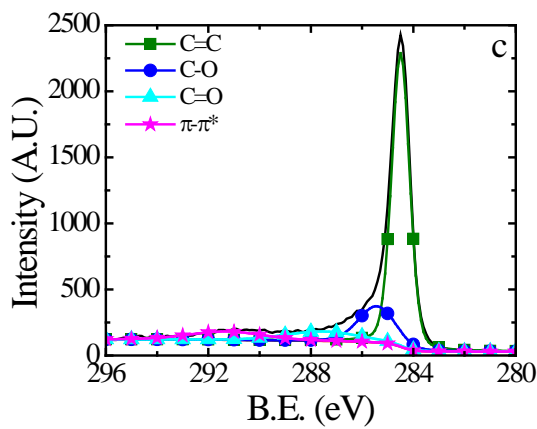
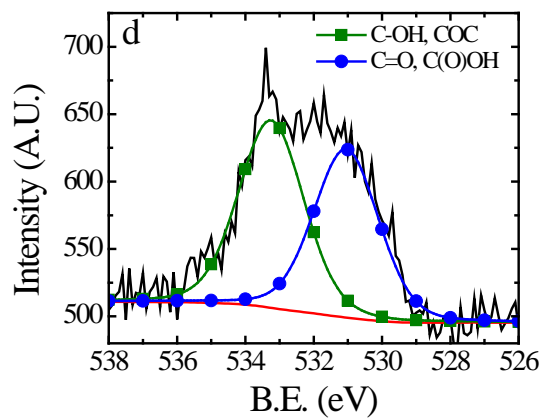

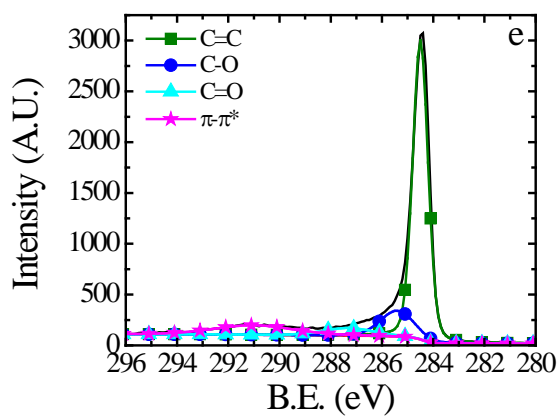
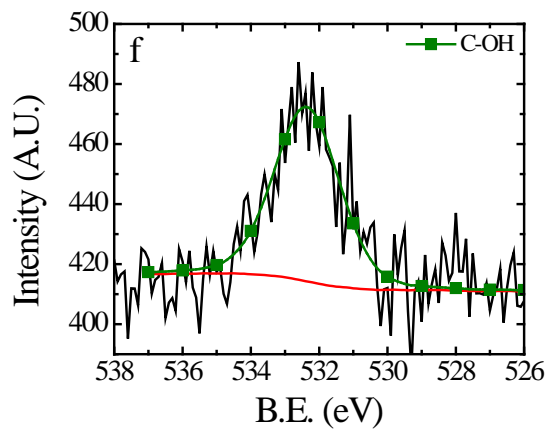
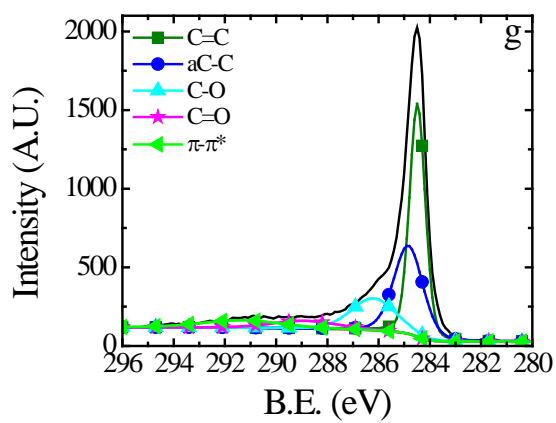
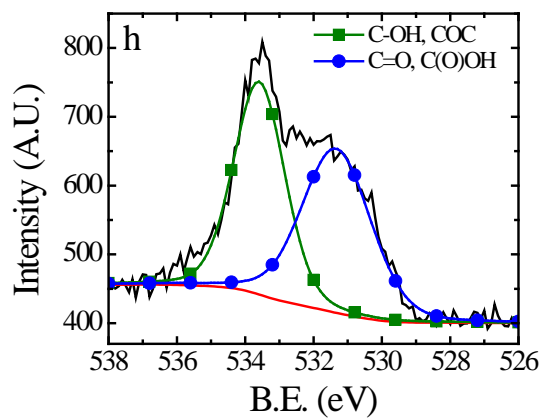
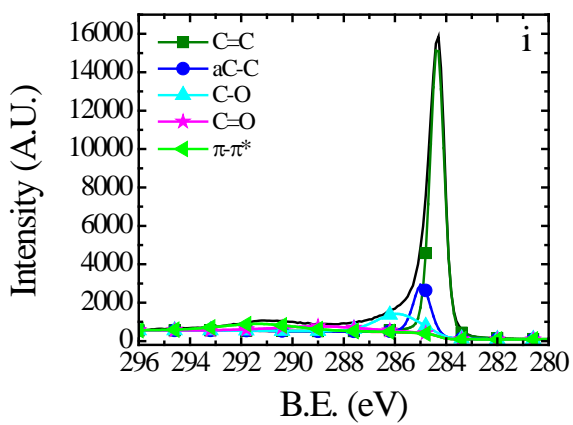
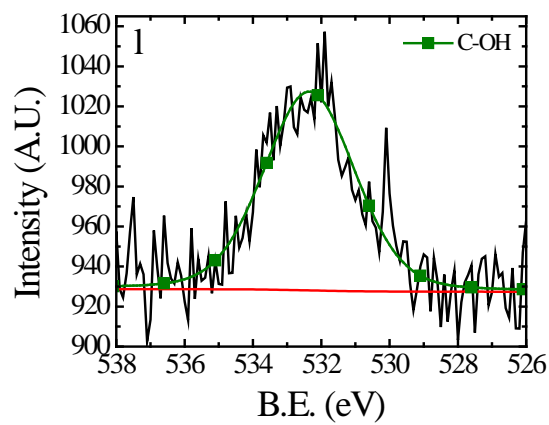

**Figure S2**. XPS curves with their deconvolution peaks for C1s and O1s of GNP (a, b), RGO (c, d), RGO_1700 (e, f), RGO-2 (g, h) and RGO-2_1700 (i, l), respectively.

XPS C1s results for GNP (**Figure** a) shows the presence of an intense and narrow sp$^2$ C-C signal coupled with a relatively intense signal due to π-π* shake-up transition which reveal a good aromaticity degree of the graphitic structure; the fitting of the C1s is completed by two weak peaks at 285.5 and 287.3 eV which, accordingly with literature, were assigned to C-O (C-OH and/or C-O-C) and C=O (C=O and/or O=C-OH) chemical bonds, respectively. Deconvolution of oxygen 1s band required the use of three peaks centered at 531.6 (double bonded oxygen), 532.4 (C-O-C) and 533.1 eV (C-OH) suggesting an higher content of double bonded respect to single bonded oxygen.

For RGO, deconvolution of C1s peak leads to similar results as those observe for GNP with carbon-oxygen bond peaks centered in the same positions. However, it is worth noting that the full width half maximum (FWHM) for sp$^2$ C-C bond of RGO is greater than that of GNP (0.84 and 0.67 for RGO and GNP, respectively), indicating an higher heterogeneity on the structural environment of carbon[7]. Moreover the intensity of the peak at 285.5 eV is higher, reflecting the oxygen calculation made by survey scans. Due to the low intensity of the oxygen 1s B.E. region, a reliable fitting was obtained with only two peaks, located at 533.2 and 531.0 eV, indicating the coexistence of single and double bonded oxygen.

After annealing of RGO, the intensity of the two peaks in the C1s region, related to carbon bonded with oxygen, decreases, as calculated with survey scans. Furthermore, the FWHM of sp$^2$ C diminishes down to 0.71 eV indicating an higher homogeneity of the graphitic structure. O1s band is very weak and can be convolved with only one Voigt centered at 532.4 eV which was related to some single bonded C-O mainly in the form of phenolic groups[3].

Fitting of C1s band for RGO-2 requires the addition of a further Lorentzian-Gaussian at 284.8 eV in order to have a more reliable deconvolved signal. The presence of an additional peak in this position was previously reported to be related to disordered carbon[7] and is in good agreement with the results obtained with Raman spectroscopy (Figure 3). As observed for RGO, even for the O1s signal of RGO-2 two partially overlapping peaks were clearly detected and a simple deconvolution was preferred, with two peaks at 531.3 eV (double bonded oxygen) and 533.5 eV (single bonded oxygen).

The XPS C1s and O1s bands of RGO-2_1700 exhibit significantly changes respect to pristine RGO-2. It is clearly visible a decrease of the intensity of all $sp^3$ C related signals respect to that of $sp^2$ carbon at 284.5 eV; as observed for RGO, even in this case the FWHM after annealing was 0.65 eV, respect to 0.72 eV measured for pristine RGO-2, revealing an increase in the homogeneity of the graphitic structure. Moreover, it is worth noting that the ratio between the $sp^2$ C-C and disordered carbon peak areas is drastically increased from ~1.4 up to ~5.0, further indicating the obtainment of a more homogeneous and ordered graphitic structure. Despite this some disordered carbon is still present in RGO-2_1700. Analysis on O1s band reveals a weak signal, which was fitted with only one Lorentzian-Gaussian centered at 532.3 eV, related to single bonded C-O mainly in the form of phenolic groups, as already observed for RGO_1700.

**Thermogravimetry**

Thermogravimetric plots for mass and derivative mass in air for GNP, RGOs and thermally annealed RGOs are reported in Figure S3

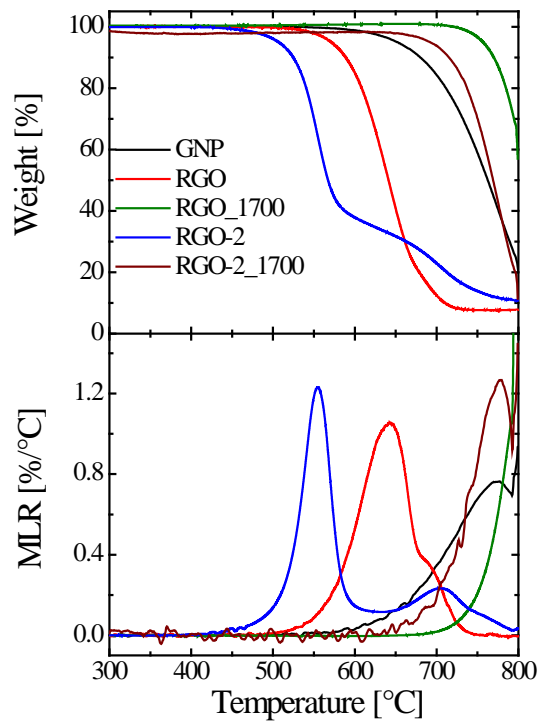

**Figure S3.** TGA curves for the different nanocomposites (Air, 10°C/min)

**Differential Scanning Calorimetry.**

DSC plots for CBT, pCBT and pCBT nanocomposites are reported in Figure S4. DSC plot on pure CBT exhibit an exothermic peak at about 80°C, which corresponds to a cold crystallization, and three separated endothermic peaks at about 125, 155 and 188°C due to the melting of different CBT oligomers with different chain length. No traces of these melting peaks are present neither in pCBT nor in nanocomposites.

Pure pCBT exhibits two partially overlapping endothermic peaks at 218.8 and 226.4°C, respectively, during heating. This is a well-known behavior for pCBT and the peak at lower temperatures is related to small crystals which melt and recrystallize to re-melt again at higher temperatures [8]. In the presence of nanoflakes more stable pCBT crystal are obtained and no clear melting/recrystallization peaks were observed.

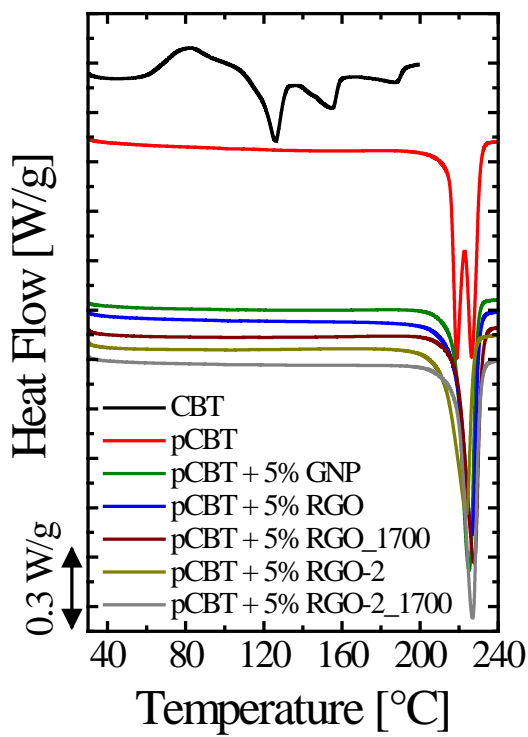

**Figure S4**. DSC curves on heating (10°C/min) for CBT and pCBT

**Table S1**. Crystallinity degree of pCBT and its nanocomposites

|  | Crystallinity degree [%] |
|---|---|
| pCBT | 41.7 |
| pCBT + 5% GNP | 40.0 |
| pCBT + 5% RGO | 38.0 |
| pCBT + 5% RGO_1700 | 39.4 |
| pCBT + 5% RGO-2 | 34.4 |
| pCBT + 5% RGO-2_1700 | 40.9 |

**Dynamo-mechanical thermal analysis**

Temperature for peak of the alfa transition in DMTA and width of the transition are reported in Table S2

**Table S2**. Tanδ peak and FWHM for pCBT and its nanocomposites

| | $T_{peak}$ [°C] | FWHM [°C] |
|---|---|---|
| pCBT | 61.5 | 41.6 |
| pCBT + 5% GNP | 64.5 | 41.8 |
| pCBT + 5% RGO | 68.4 | 52.6 |
| pCBT + 5% RGO_1700 | 64.9 | 53.8 |
| pCBT + 5% RGO-2 | 61.3 | 42.2 |
| pCBT + 5% RGO-2_1700 | 67.5 | 42.0 |